\documentclass[useAMS, usenatbib]{mn2e}
\usepackage{aas_macros}
\usepackage{amsfonts}
\usepackage{amsmath}
\usepackage{graphicx}

\newcommand{\pa}{\varphi_A}
\newcommand{\pb}{\varphi_B}
\newcommand{\pc}{\varphi_C}
\newcommand{\bs}{\boldsymbol}
\newcommand{\av}{\overline}

\title[Phase Errors in Diffraction-Limited Imaging]{Phase Errors in Diffraction-Limited Imaging: Contrast Limits for Sparse Aperture Masking}

\author[M.J. Ireland]{M.J. Ireland\thanks{mireland@aao.gov.au} \\
$^1$Australian Astronomical Observatory,  PO Box 296, Epping, NSW 1710, Australia \\
$^2$Department of Physics and Astronomy, Macquarie University, NSW 2109, Australia\\
$^3$Macquarie University Research Centre in Astronomy, Astrophysics \& Astrophotonics
}

\begin{document}

%ApJ following 4 lines
%\author{Ireland, M.J.}
%\affil{Australian Astronomical Observatory, PO Box 296, Epping NSW 1710, Australia}
%\affil{Department of Physics \& Astronomy, Macquarie University, NSW 2109, Australia} 
%\affil{Macquarie University Research Centre in Astronomy, Astrophysics \& Astrophotonics}

%MNRAS following 2 lines.
\pagerange{\pageref{firstpage}--\pageref{lastpage}} \pubyear{2013}
\maketitle

\label{firstpage}

\begin{abstract}
Bispectrum phase, closure phase and their generalisation to kernel-phase
are all independent of pupil-plane phase errors to first-order. This property, when used with Sparse Aperture Masking 
(SAM) behind adaptive optics, has been used recently in high-contrast observations
at or inside the formal diffraction limit of large telescopes. Finding the limitations to these
techniques requires an understanding of spatial and temporal third-order phase effects, as well as effects
such as 
time-variable dispersion when coupled with the non-zero bandwidths in real observations.
In this paper, formulae describing many of these errors are developed, so that a comparison can be made to
fundamental noise processes of photon- and background-noise. 
I show that the current generation of aperture-masking 
observations of young solar-type stars, taken carefully in excellent observing conditions, are consistent with being 
limited by temporal phase noise and 
photon noise. This has relevance for plans to combine pupil-remapping with spatial filtering. 
Finally, I describe calibration strategies for kernel-phase, including the optimised calibrator weighting
as used for LkCa15, and the restricted kernel-phase {\small POISE} technique that avoids explicit dependence on
calibrators.
\end{abstract}

\section{Introduction} 
\label{secIntro}
The concepts of closure-phase, bispectrum phase \citep[e.g.][]{Hofmann93}, self-calibration and now kernel-phase \citep{Martinache10}
are well-known as techniques that cancel out many instrumental effects due to pupil-plane phase errors. 
Despite the very long history of aperture-masking with a focus on fringe visibility amplitude \citep{Fizeau1868,Michelson1891,Schwarzschild1896},
it was the use of closure-phase that first enabled image-reconstruction from this 
technique \citep{Baldwin86} as well
as recent efforts in high-contrast imaging \citep[e.g.][]{Lloyd06, Kraus12}. 

A simple explanation of closure-phase comes from a counting argument. From an interferometer with $M$ (sub)-apertures, 
the complex visibilities can be independently measured on each of the $M(M-1)/2$ baselines consisting of each pair of 
(sub)-apertures. An optical aberration consisting of a piston on each of the (sub)-apertures amounts to $M-1$ degrees of 
freedom in the phase differences, leaving $(M-1)(M-2)/2$ additional measured quantities, which are the linearly-independent
set of closure-phases. A set of observables which are independent of pupil-plane phase form an ideal starting point for precise
model-fitting and imaging at the diffraction-limit. This argument applies to both redundant and non-redundant pupil geometries, as realised by 
\citet{Martinache10}.  But if phase errors on a pupil are large, a redundant pupil configuration is at a disadvantage, because the pairs of
pupil locations that form any given Fourier component may add out-of-phase and destructively interfere. In the case of
observations taken behind adaptive optics, the choice of one technique over the other is not obvious.

In this paper, I will outline the causes of contrast limitations in the aperture-masking interferometry and kernel-phase techniques, 
and methods to maximise contrast. In Section~\ref{sectCauses} the main causes of kernel-phase errors will be outlined. In 
Section~\ref{sectCorrelations} I will describe why the statistical correlations between closure-phases mean that kernel-phases
are preferred as a primary observable, and will compare the contrast limits achievable by different pupil geometries. In Section~\ref{sectNearestNeighbour} I will describe standard closure-phase calibration and its limitations, in 
Section~\ref{sectGOPI} I will describe the calibration strategy as used in \citet{Kraus12} to maximise contrast in aperture-masking interferometry observations, 
and in Section~\ref{sectPOISE} I will describe the simpler {\small POISE} calibration strategy.
%Phase Observationally Independent of Systematics Errors
In Section~\ref{sectConclusions} I will conclude and 
outline the key areas where further research is needed.

\subsection{Kernel-Phase}
\label{kerphaseIntro}

The definition of Kernel-phase as used in this paper will be slightly simplified from the definition of \citet{Martinache10}, as we will avoid the use of the ``redundancy'' matrix $\bs{R}$. To first-order in pupil-plane phase (i.e. with a nearly-flat wavefront), we can write the observed phase $\Phi_m$ in the Fourier transform of an image as:

\begin{equation}
\Phi_m = \bs{A} \cdot \varphi+ \Phi_o,
\end{equation}

where $\varphi$ is the pupil-plane phase and $\Phi_o$ is the phase of the Fourier transform of the object.  
These are represented as vectors where each vector element is one discrete point in the model pupil plane or the image discrete
Fourier transform.
The matrix $\bs{A}$ encodes the information about which parts of the pupil form each Fourier component. For example, a non-redundant baseline formed by two discrete pupil components only would have a +1 and -1 in that row of $\bs{A}$, with all other elements taking the value 0.  
This matrix is described in detail in \citet{Martinache10}.
%Back to normal (minor wording change only)
Using singular value decomposition, we then find a matrix $\bs{K}$, the Kernel of $\bs{A}$, such that $\bs{K}\cdot \bs{A}=0$. 
By choosing $\bs{K}$ such that its number of non-zero rows is equal to its rand, this matrix enables us to 
project the Fourier phases onto a subspace, which we will call the {\em Kernel-phases} $\theta$ by $\theta = \bs{K}\cdot\Phi$. On this subspace, the observables are not affected by pupil-plane phase errors at first-order:

%\begin{eqnarray}
%\theta_m &=& \bs{K} \cdot \Phi_m \nonumber \\
% &=& (\bs{K} \cdot \bs{A}) \cdot \varphi + \bs{K} \cdot \Phi_o \nonumber \\
% &=& \bs{K} \cdot \Phi_o
%\end{eqnarray}
\begin{align}
\theta_m &= \bs{K} \cdot \Phi_m \nonumber \\
 &= (\bs{K} \cdot \bs{A}) \cdot \varphi + \bs{K} \cdot \Phi_o \nonumber \\
 &= \bs{K} \cdot \Phi_o
\end{align}

A model of the object can therefore be directly compared to the observed Kernel-phases by computing the Fourier transform and multiplying by the matrix $\bs{K}$. 
For all reasonable 2-dimensional pupils, the rank of $\bs{A}$ is at least half the length of $\Phi_o$, meaning that at least half the object Fourier-phase information is preserved when transforming from Fourier-phase to kernel-phase.

\section{Causes of Kernel-Phase Errors}
\label{sectCauses}

There are three broad classes of kernel-phase errors: those that vary rapidly, approximating white noise in a sequence of exposures (random errors), those that are static throughout an observing run and can therefore be calibrated by observation 
of unresolved calibrator stars (static errors) and those which vary from one target to another (calibration errors). Calibration errors include quasi-static errors with a time variability measured in minutes or hours, as well as errors that depend on e.g. the sky position or the spectrum of the source observed. The goal of any combination of observing technique and analysis strategy is to both minimise the random errors, and to develop a calibration strategy where residual calibration errors are smaller than typical random errors. The following sections include error causes that could manifest themselves as one or several of these error classes.

\subsection{General Pupil-Plane Phase Errors}
\label{sectGeneral}

\begin{figure}
\includegraphics[width=0.3\textwidth]{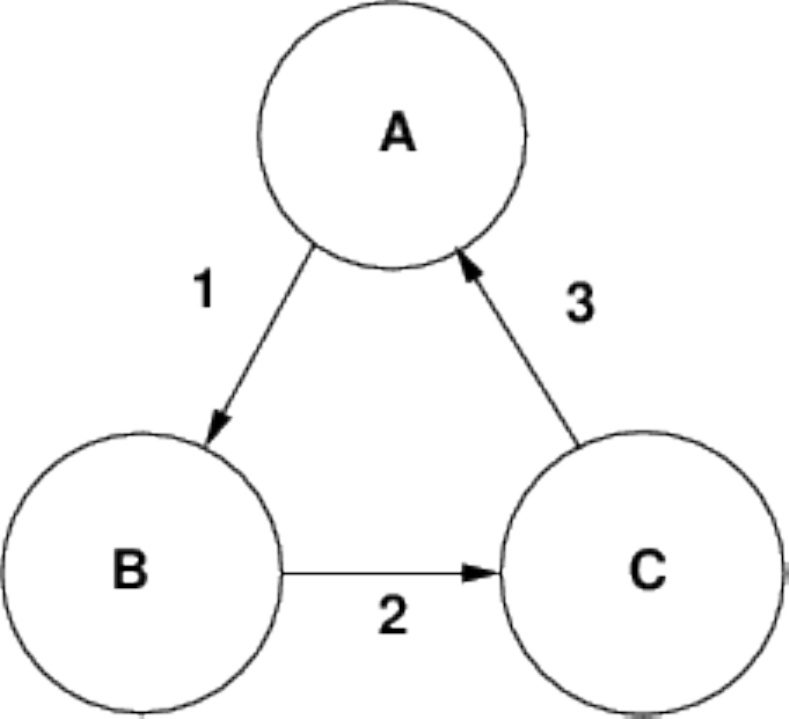}
\caption{An abstract representation of closure-phases formed by baselines 1, 2 and 3, in turn formed by 
congruent apertures $A$, $B$ and $C$.}
\label{figABC}
\end{figure}

We will examine first an abstract representation of pupil-plane phase errors that could cause random, calibration or static errors. We consider a closing triangle containing apertures $A$, $B$ and $C$, as depicted in Figure~\ref{figABC}. Each aperture has the same size and shape, and each baseline $1\equiv A\rightarrow B$, $2\equiv B\rightarrow C$ and $3\equiv C\rightarrow A$ has data taken at the same time. That is, there are equivalent coordinate systems describing apertures $A$, $B$ and $C$, centered on each aperture. This means that the visibility on each baseline is formed by the incoherent integral of visibilities arising from common spatio-temporal coordinates in sub-apertures $A$, $B$ and $C$.

We will assign the symbols $\pa$, $\pb$ and $\pc$ to the phase in sub-apertures $A$, $B$ and $C$, the symbols $\Phi_1$, $\Phi_2$ and $\Phi_3$ to the phase on baselines 1, 2 and 3 respectively, and will neglect amplitude variations (i.e. scintillation). 
The complex visibilities are then formed by:

\begin{align}
V_1 &= \av{ \exp{i (\pb - \pa)} }\nonumber \\
V_2 &= \av{ \exp{i (\pc - \pb)} }\nonumber \\
V_3 &= \av{ \exp{i (\pa - \pc)} },
\end{align}

where the bar represents an average over the spatio-temporal co-ordinates corresponding to each aperture. This can be expanded to third-order in phase to:

\begin{equation}
V_1 \approx 1 + i \av{(\pb - \pa)} - \frac{1}{2}\av{(\pb-\pa)^2} - \frac{i}{6}\av{(\pb-\pa)^3},
\label{eqnThirdOrderV}
\end{equation}

with similar expressions for $V_2$ and $V_3$. The bispectrum is given by the product of these three visibilities, which can be again expanded to third-order in phase:

\begin{equation}
b_{ABC} = V_1 V_2 V_3
\label{eqnBispect}
\end{equation}

\begin{align}
\Re(b_{ABC}) &\approx 1 - \frac{1}{2}[\av{(\pb' - \pa')^2} + \av{(\pc' - \pb')^2} \nonumber \\
 & ~~~ + \av{(\pa'-\pc')^2}] \label{eqnRealBispect} \\
\Im(b_{ABC}) &\approx  -\frac{1}{6}[\av{(\pb' - \pa')^3} + \av{(\pc' - \pb')^3} \nonumber \\
 & ~~~ + \av{(\pa'-\pc')^3}] \label{eqnImBispect},
\end{align}

where we have considerably simplified the expansion by introducing the piston-corrected phases: 

\begin{align}
\pa' &= \pa - \av{\pa} \label{eqnPistonA} \\
\pb' &= \pb - \av{\pb} \label{eqnPistonB} \\
\pb' &= \pb - \av{\pb} \label{eqnPistonC}
\end{align}

A more complete derivation of this expansion is given in Appendix~A. The closure phase $\phi_{\rm cp} = \Phi_1 + \Phi_2 + \Phi_3$ is then most simply approximated by taking the leading terms in the real (0th order) and imaginary (3rd order) components of the bispectrum, giving $\phi_{\rm cp} = \Im(b_{ABC})$.

It is also worthwhile briefly considering the effects of averaging the visibilities for baselines 1, 2 and 3 over different spaces. This could be caused by differing sub-aperture shapes in conventional aperture-masking interferometry (amounting to non-closing triangles), or by disjoint integration times
as found in other forms of interferometry. In this case, the leading terms in the closure-phase errors become first-order rather than third order in pupil plane phase. Clearly, this is something to be avoided at considerable effort in the case of high-contrast aperture-masking. The pupil ``shape'' can also be thought of as the pupil-plane amplitude within each sub-aperture. Where amplitude errors are taken into account, these closure-phase errors then become second-order, i.e. first-order in phase and first-order in amplitude, and could plausibly be the leading term.

%XXX A figure here is easy, with an additional part to Figure 1.

\subsection{Temporal Phase Errors}
\label{sectTemporalErrors}

Our first application of Equation~\ref{eqnImBispect} to closure-phase errors is rapid temporal effects, which cause a random kernel-phase error.  There are two key regimes that temporal errors operate in behind an AO system. Either exposure-times are comparable to or shorter than the inverse of the AO system bandwidth (the short-exposure regime) or exposure times are significantly longer than these timescales (the long-exposure regime). Given typical coherence times at $\sim$2.2 microns or shorter wavelengths of $<$50\,ms, and typical AO system bandwidths in the range 10-100\,Hz, exposure times longer than $\sim$100\,ms in the near-infrared are in the long-exposure regime.

In the long-exposure regime, we can make the approximation that piston noise is white up to some cutoff frequency $f_c$. This is not very unrealistic, because in the frozen turbulence approximation, the atmosphere has an amplitude spectrum proportional to $f^{-5/6}$, while the error signal from a Proportional-Integral-Differential (PID) controller in the mid-frequency range where the proportional term dominates gives residual errors proportional to the input signal amplitude multiplied by the frequency $f$. This gives a resultant error amplitude proportional to $f^{1/6}$, up to the servo loop cutoff. At this cutoff, three independent phenomena all tend to cut-off the error spectrum rapidly: the $f^{-5/6}$ atmospheric amplitude spectrum, the rapidly lowering gain of the servo approaching its Nyquist sampling frequency, and effects of spatial filtering.

We will now make a second set of approximations by assuming that the phase piston on each sub-aperture making a closing-triangle is un-correlated and has  identical phase noise $\sigma_{\varphi}$. This may not be reasonable for some AO systems (e.g. if tip/tilt errors dominate due to tip/tilt mirror bandwidth) but as this depends on reconstructor and wavefront sensor details, it is a good first approximation.

An exposure of total time $T$ can then be split into $f_c T$ sub-exposures, each of which has independent phase noise, so that in each exposure we have pupil-plane sub-aperture piston phases given by normal distributions:

\begin{align}
\pa &\sim \mathcal{N}(0,\sigma_{\varphi}) \\
\pb &\sim \mathcal{N}(0,\sigma_{\varphi}) \\
\pc &\sim \mathcal{N}(0,\sigma_{\varphi}).
\end{align}

Applying Equation~\ref{eqnImBispect} to this phase noise distribution for $f_c T >> 1$ gives the standard deviation of closure-phase (see Appendix~B for a derivation):

\begin{equation}
\sigma(\phi_{\rm cp, temporal}) = \sigma_{\varphi}^{3} \sqrt{3 / f_c T} ~{\rm rad}.
\label{eqnTemporalErrors}
\end{equation}

In the short exposure regime, we are dominated by atmospheric piston, as in the case with aperture-masking interferometry without adaptive optics (e.g. \citealt{Tuthill00}). In this regime, for typical exposure times $\Delta t$ less than $\sim$20\,ms at a 2.2\,$\mu$m wavelength, or $\sim$50\,ms at 4\,$\mu$m wavelength without adaptive optics or fringe-tracking, we can still consider phase errors at third-order with reasonable accuracy. By evaluating Equation~\ref{eqnImBispect} numerically based on Kolmogorov turbulence, we arrive at:

\begin{equation}
\sigma(\phi_{\rm cp, temporal}) = 0.0177 (\frac{\Delta t}{t_0})^{15/6}~{\rm rad.},
\end{equation}

which is valid for $\Delta t  \la t_0$.  This kind of relationship also has relevance to long-baseline interferometry in the case of measurements where visibilities are measured simultaneously. Examples of this are MIRC \citep{Monnier06} or PAVO \citep{Ireland08c}  at the CHARA array. This relationship does not apply to scanning beam combiners, where fringes can be recorded non-simultaneously depending on group delay tracking accuracy.

\subsection{Spatial Closure-Phase Errors}
\label{sectSpatial}

In this section, we will examine how wavefront phase corrugations affect closure- or kernel- phases, occurring as random, calibration and static errors. Calibration errors occur when there are slowly time-variable spatial aberrations 
(often called {\it quasi-static} speckles). 
To most easily compare kernel phase to closure-phase, we adopt a factor of $1/\sqrt{3}$ scaling to the closure-phase, so that adding the three baseline phases is equivalent to multiplying by a unit vector (e.g. one of the orthonormal columns of the matrix $\mathbf{V}$ from \citealt{Martinache10}). 

\begin{figure*}
\includegraphics[width=\textwidth]{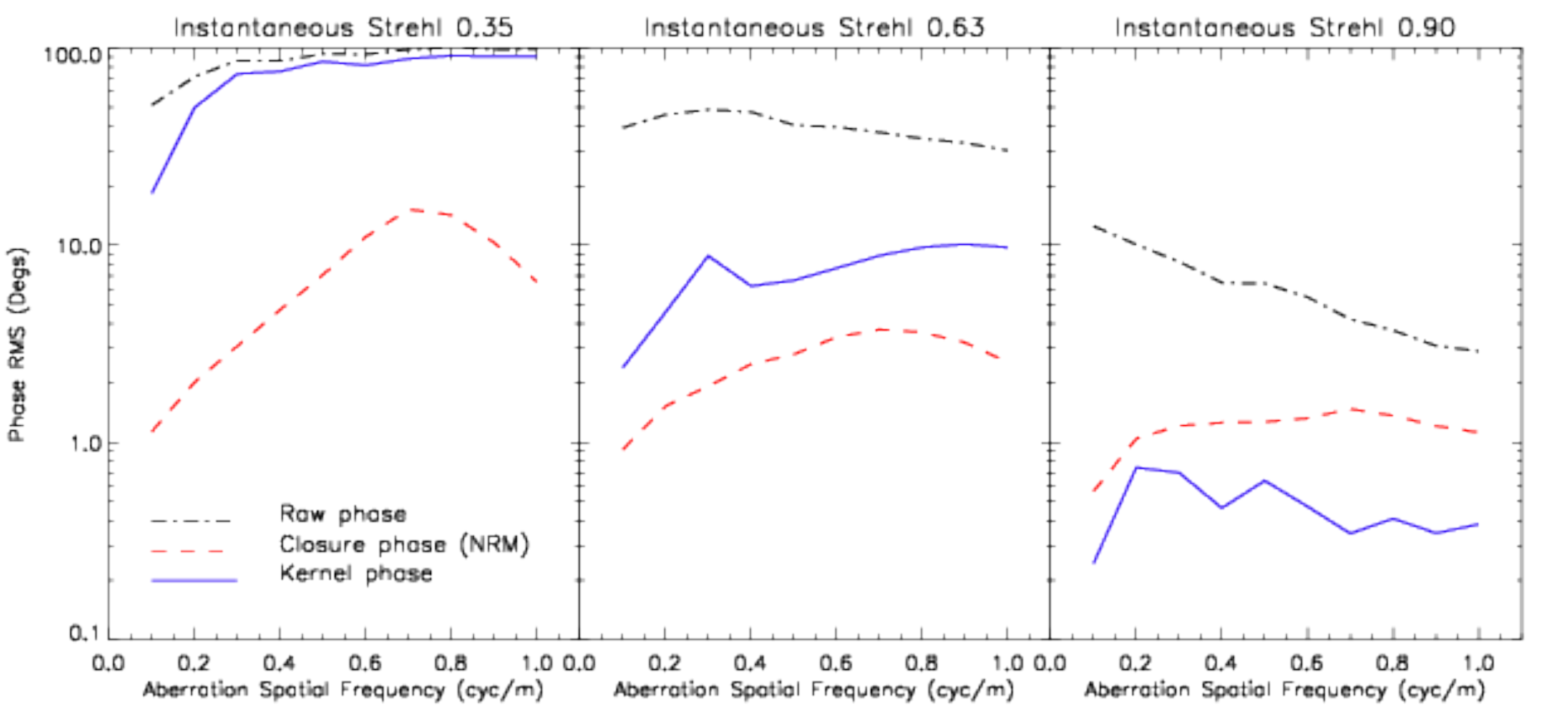}
\caption{The effect of RMS pupil-plane phase errors of 1 radian (left), 0.7 radians (centre) and 0.35 radians (right) on raw
aperture-masking Fourier phase (black dot-dashed), full-pupil Kernel-phase (blue) and aperture-masking closure-phase (red) scaled by a factor of $1/\sqrt{3}$ as described in the text.  The pupil geometries are the Keck non-redundant 9-hole mask and the full 
Keck pupil.}
\label{figPhaseErrors}
\end{figure*}

Figure~\ref{figPhaseErrors} shows a comparison between simulated sparse aperture-masking and kernel-phase data analysis for a variety of aberration spatial frequencies and aberration amplitudes. For each amplitude and spatial frequency, the position angle of a sinusoidal aberration was randomly varied and the overall RMS kernel-phase computed. It can be seen that although both kernel phase and closure-phase appear equivalent to first-order, they have quite different responses to high-order pupil plane errors. The spatial filtering of an aperture-mask means that it can be effectively used at much lower instantaneous Strehl ratios than unobstructed-pupil kernel-phase, but in a high-Strehl regime, kernel-phase is in principle superior. For the 0.35 radians RMS phase error case (right hand figure), Equation~\ref{eqnImBispect} predicts closure-phases approximately 2 times lower than the simulation, possibly due to Fourier sampling and windowing effects in the sparse aperture-masking pipeline used, and possibly due to effects higher than 3rd order in pupil-plane phase. For very high instantaneous Strehls, kernel phase in both geometries is expected to scale as the cube of the pupil-plane phase error, which is $(1-S)^{3/2}$ in the Mar\'echal approximation.

A comparison between imaging with an unobstructed aperture and with sparse aperture masks is complicated somewhat by the ability to window data, which smooths over high spatial frequency aberrations. This gives a further advantage in-principle to an unobstructed aperture or a mask with large holes where the interferogram has a relatively small spatial extent. An example of a regime where fine spatial scale aberrations may dominate phase errors post-calibration is when aberrated pupil-plane elements or masks shift due to flexure effects.

\subsection{Flat Field Errors}
\label{sectFlatField}

In sparse aperture masking, many pixels are used to record fringes from objects with intrinsically small spatial extents. 
If target and calibrator objects are not acquired on the same pixels, then the effect of flat field errors is to add random phase errors across the Fourier plane. These random errors are only static if alignment is perfect between target and calibrator star observations -- otherwise flat field errors become a calibration error. A flat field error can be modelled as multiplication in the image plane by a function that is 1.0 everywhere plus white-noise with standard deviation $\sigma_F$. A typical value of $\sigma_F$ is 10$^{-3}$, arising from a series of flat field exposures with a total of $10^6$ photo-electrons per pixel. Multiplication by this flat is equivalent to convolution in the Fourier domain, which spreads the power from the zero and near-zero spatial frequency components over the full Fourier plane. Clearly phase errors will then be proportional to $\sigma_F$ and inversely proportional to visibility. Numerical simulations give the following relationship for closure-phase in sparse aperture-masking observations:

\begin{equation}
 \sigma(\phi_{\rm cp, photon}) \la 0.3 \frac{\sigma_F}{V}~{~rad.},
\end{equation}

where $V$ the fringe visibility, referenced to a perfect Strehl interferogram of a point source. The constant of $\sim$0.3 varies between approximately 0.2 and 0.3 for different bandpass filters and aperture masks. To ensure that these errors are less than $10^{-3}$ radians with typical visibilities of 0.3, we need $\sigma_F < 10^{-3}$, meaning at least $10^6$ photons per pixel recorded when taking flat fields.

\subsection{Bad Pixels}
\label{secBadPixels}

The existence of bad pixels on an imaging array can often destroy sensitivity in traditional imaging over a small portion of the field of view. 
Like flat field errors, incorrectly accounting for bad pixels can cause significant calibration errors. By spreading the information over many pixels, it may seem that at first glance bad pixels would always do significant harm to the information content in aperture-masking observations. However, the limited Fourier support of this kind of observation, as long as it is better than Nyquist sampled, means that bad pixels can be very effectively corrected. In simulations, the algorithm below has proved effective at contrasts beyond 10$^6$ for arrays far worse than those found at telescopes where aperture-masks are installed,
meaning that if properly corrected, bad pixels are not a cause of kernel-phase errors.

The principle of this bad pixel correction algorithm is to assign the values to the bad pixels so that the power in the Fourier domain outside the region of support permitted by the pupil geometry is minimised. We will call this region of the Fourier plane the zero region $Z$. We can turn this problem into a linear one by realising that the Fourier components corresponding to the set of bad pixel coordinates $\mathbf{x}_b$ forms a subspace of $Z$, and we can find a vector of bad pixel offsets $\mathbf{b}$ to subtract so that the image Fourier transform on this subspace is identically zero.

The first step in this process is to create the matrix $\mathbf{B_z}$ which maps the bad pixel values onto $Z$. The measured values $f_Z$ in the Fourier plane region $Z$ are then modelled as:

\begin{equation}
f_Z = \mathbf{B}_Z \cdot \mathbf{b}  + \epsilon_Z,
\end{equation}

with $\epsilon_Z$ being the remaining Fourier-plane noise. The bad pixel adjustments $\mathbf{b}$ are then found using the Moore-Penrose pseudo-inverse of $\mathbf{B}$:

\begin{align}
\mathbf{b} &= \mathbf{B}_Z^+ \cdot f_Z \label{eqnBadPixelCorrection} \\
 &= (\mathbf{B}_Z^* \cdot \mathbf{B}_Z)^{-1} \cdot \mathbf{B_Z}^* \cdot f_Z \label{eqnDodgyInverse}
\end{align}

The Moore-Penrose pseudo-inverse can also be found by other methods such as singular-value-decomposition rather than direct computation of an inverse as in Equation~\ref{eqnDodgyInverse}, but this method suffices for a relatively small number of bad pixels. Although this algorithm is very quick (the matrix $\mathbf{B}^+$ is pre-computed), the bad pixel correction Equation~\ref{eqnBadPixelCorrection} does have to be applied for every frame, with the computed values $\mathbf{b}$ subtracted off each frame. It can also be used to correct for saturated pixels at the core of a PSF,
pixels affected by transient events such as cosmic rays, or an acquisition error where a small portion of the interferogram is truncated by the detector edge.

%\subsection{Fringe-Scanning Errors}
%\label{sectScanning}
%
%Many of the above sections have been mostly applicable to aperture-masking interferometry. In long-baseline interferometry, there %are additional effects that can limit closure-phase precision. One of the worst offenders is the effective non-simultaneity

\subsection{Dispersion and Wavelength-Dependent Phase Errors}

Kernel-phase observations are often made in a broad-band filter where different wavelengths are affected by both the atmosphere and optics in different ways. This causes a static kernel-phase error, which can become a calibration error unless observing conditions and spectrum are matched between target and calibrator observations. A general analysis of these errors is particularly difficult and beyond the scope of this paper, because the definition of kernel-phase is inherently monochromatic. 
However, we can put some limits on when this effect might become important, and the order of magnitude of the effect. We write the air refractive index difference of $\Delta n$ between the blue and red edges of a filter, and the spectral difference between a target and calibrator is $\Delta F$ covering a fraction $f$ of the bandpass. Assume both objects are observed at the same airmass. The image Fourier-plane phase error arising from this 
difference is:

\begin{align}
\Delta \varphi &\approx  2\pi \Delta F  f  \Delta \alpha B_{\rm max}/ \lambda_{\rm mean},
\end{align}

where the change in angle on the sky between long and short wavelength part of the filter is:

\begin{align}
\Delta \alpha =  \Delta n \tan(z) .
\end{align}

Here $z$ is the zenith distance angle, and this formula is only value for air masses less than approximately 3. The kernel-phase signature of this dispersion effect is very similar to that of a close companion of separation $\Delta \alpha$ and magnitude difference $f \Delta F$. For values of $\Delta \alpha$ greater than about $0.5 \lambda_{\rm mean}/B_{\rm max}$, the kernel-phase error $\Delta \theta$ is of the same magnitude as $\Delta \varphi$, and for smaller values of  $\Delta \alpha$, the kernel-phase error goes as $(\Delta \alpha B_{\rm max}/ \lambda_{\rm mean})^3$ \citep[e.g. see Equation 5 of][]{leBouquin12}. As an example, observing in the full H-band with a zenith angle of 45 degrees from an altitude of 2600\,m gives $\Delta \alpha=31$\,milli-arcsec, which is larger than $0.5 \lambda_{\rm mean}/ B_{\rm max}$ for $B_{\rm max} = 8$\,m. A 10\% difference in the spectrum over the long-wavelength 10\% of the H bandpass would then give $\Delta \theta \approx \Delta \varphi \approx 0.01$ radians. 

The effect of observing at different airmass is much more complex, because for flat spectra, dispersion does not give a non-zero kernel-phase. In general, it may be a non-linear interaction between pupil-plane aberrations and dispersion that dominate the calibration errors.

\subsection{Photon, Background and Readout Noise}

Finally, we consider the fundamental limitation of random errors caused by photon, background and readout noise. Where the fringe visibility is $V$, the total number of photons collected in an interferogram is $N_p$, the number of background photons $N_b$ and the number of holes in the aperture mask $N_h$, the closure-phase error due to photon (shot) noise is:

\begin{equation}
\sigma(\phi_{\rm cp, photon}) =  \frac{N_h }{N_p V} \sqrt{1.5(N_p + N_b + n_p \sigma^2_{\rm ro})}. \label{eqnPhotonNoise}
\end{equation}

The factor of $\sqrt{1.5}$ includes a factor of $\sqrt{3}$ due to photon noise from three independent baselines making up the closure-phase, as well as a factor of $\sqrt{1/2}$ due to the shot noise power at any non-zero spatial frequency being split equally between the real and imaginary parts. The readout noise in photon units is  $\sigma_{\rm ro}$ and the number of pixels $n_p$. The effect of both readout and background noise is affected by the size of the window function used prior to making the Fourier transform to compute the visibilities, and this effect can be minimised if fringes are directly fit to the data (e.g. the {\small SAMP} pipeline of \citet{Lacour11}).

\subsection{Dominant Error Terms}

The most common kind of kernel-phase data taken so far has been sparse aperture-masking behind natural guide star adaptive optics, particularly at 1.5-2.4 micron wavelengths, so we will consider this regime first. We will also consider that adequate flat-fields have been taken and bad pixels properly corrected. The adaptive optics system only locks when there are at least $\sim$100 visible photons per Shack-Hartmann lenslet in $\sim$0.01s, or $\sim$10$^6$ photons  in 100\,s. With a similar near-infrared and visible photon rate, and a similar masking sub-aperture size to a Shack-Hartmann lenslet size, Equation~\ref{eqnPhotonNoise} would predict a $\sim$0.4 degree photon-limited closure-phase uncertainty for a 100\,s integration and a 9-hole aperture mask. 

We can use Equation~\ref{eqnTemporalErrors} to predict the effect of temporal phase errors: in particularly good seeing, $\sigma(\varphi)$ could be as low as 0.3 radians (giving a temporal phase-noise limited Strehl of $\sim$0.9) and $f_c$ could have a value of 10\,Hz. This would give a temporal phase-noise component to closure-phase uncertainty of $\sim$0.1 degrees. Perhaps not surprisingly given how much light an aperture-mask blocks, photon noise would dominate in this regime. However, for less than ideal seeing conditions and targets which are brighter in the infrared, the temporal phase noise dominates over photon noise. A characteristic ``typical seeing" predicted closure-phase error for 0.5 radians RMS pupil-plane phase error is 0.5 degrees for a 100\,integration.

% Some attempted tests of the wavelength dependence of closure-phases are check_cp_scaling.script and
% (not completed) check_cp_scaling.
%Jcont and Hcont: 100404É sort-of.
%CH4S and Kp: 101128 (DI Tau). At 1.59 and 2.13 microns, the raw closure-phase errors should be 2.4 times worse.

The closure-phase uncertainties predicted here are similar to the typical closure-phase uncertainties computed from the standard error of the mean of individual observation sets in survey papers such as \citet{Kraus08}. However, it is certainly true that the residuals when subtracting closure-phases from two point-sources are not always statistically consistent with these standard errors.
This kind of residual is often called a {\em calibration error}, where the non-zero closure phases described in Section~\ref{sectSpatial} are not fully corrected by observations of a calibrator star. Typical uncalibrated closure phases from the Keck 9 hole aperture mask are 3.5 degrees in H and K bands (CH4S and Kp filters), and 7 degrees in L band (Lp filter). These non-zero closure phases are consistent with having quasi-static spatial aberrations of $\sim$0.5 radians amplitude in the CH4S and Kp filters (e.g. Figure~\ref{figPhaseErrors}) and atmospheric dispersion in the Lp filter (Section~\ref{figNegCalibrator}). A small change in the cause of these non-zero closure phases causes miscalibrations that can be larger than the temporal (sub-aperture piston) phase and photon noise effects.

\section{Closure-Phase Correlations}
\label{sectCorrelations}

One of the more confusing aspects of aperture-masking data analysis is knowing what to do with a linearly dependent set of closure-phases. As described in \citet{Kulkarni89}, these phases may be linearly independent in the case of very low signal-to-noise per exposure when the bispectrum is averaged, but in the high signal-to-noise limit considered here, with $M$ non-redundant sub-apertures, there are $M(M-1)(M-2)/6$ closure-phases but only $(M-1)(M-2)/2$ linearly independent closure-phases.  A redundant aperture has an even higher degree of correlation of the bispectrum phases. 

Simply choosing an arbitrary independent set of closure-phases for the purpose of modelling is not possible without a full consideration of the covariance matrix. If one considers only the simplest forms of closure-phase errors, namely that due to readout-noise, then the problem of modelling the covariance matrix is not difficult. However, there are many other kinds of errors that can cause correlations between closure-phase errors.

Previous work has either gone to great lengths to diagonalise the measured covariance matrix of closure-phase \citep[e.g.][]{Kraus08} or has made an approximate scaling of fitting errors to account for the closure-phase correlations \citep[e.g.][]{Hinkley11}. The difficulty in any approach based on real data is that the sample covariance matrix must be modelled, and can not in general be measured completely from the data. The reason for this is that where there are fewer data frames taken than independent closure-phases, the sample covariance matrix is necessarily singular.

These difficulties are all avoided if rather than considering closure-phases as a primary observable, the linear combinations that make the kernel-phases are seen as the primary observables. This has added benefits of being able to extend the aperture-mask technique to considering baselines within each sub-aperture (consequently extending the usable field of view) and using the same language for all adaptive optics image analysis that is independent of pupil-plane phase to first order.

Of course, there are many different ways to form a set of kernel-phases from a set of closure-phases, or indeed a linearly independent set of kernel-phases. \citet{Martinache10} suggested that kernel phases should be constructed so that only orthonormal linear combinations of Fourier phase are considered. However, this does not guarantee statistical independence. In the simplest case of a centrally-concentrated image limited by photon-noise, the spatial concentration of the image variance means that neighbouring Fourier components have highly correlated phase errors. This amounts to a contrast loss when considering $n$-sigma excursions of kernel-phase, because just like aperture-masking, the kernel-phase technique as described by \cite{Martinache10} has a nearly flat contrast limit curve beyond separations of $\sim \lambda/D$. However, standard imaging can have increasing contrasts as separations increase beyond the PSF centre. This apparent loss in sensitivity can be regained by properly considering the correlation between Fourier phases, as shown below.

\begin{figure}
\includegraphics[width=\columnwidth]{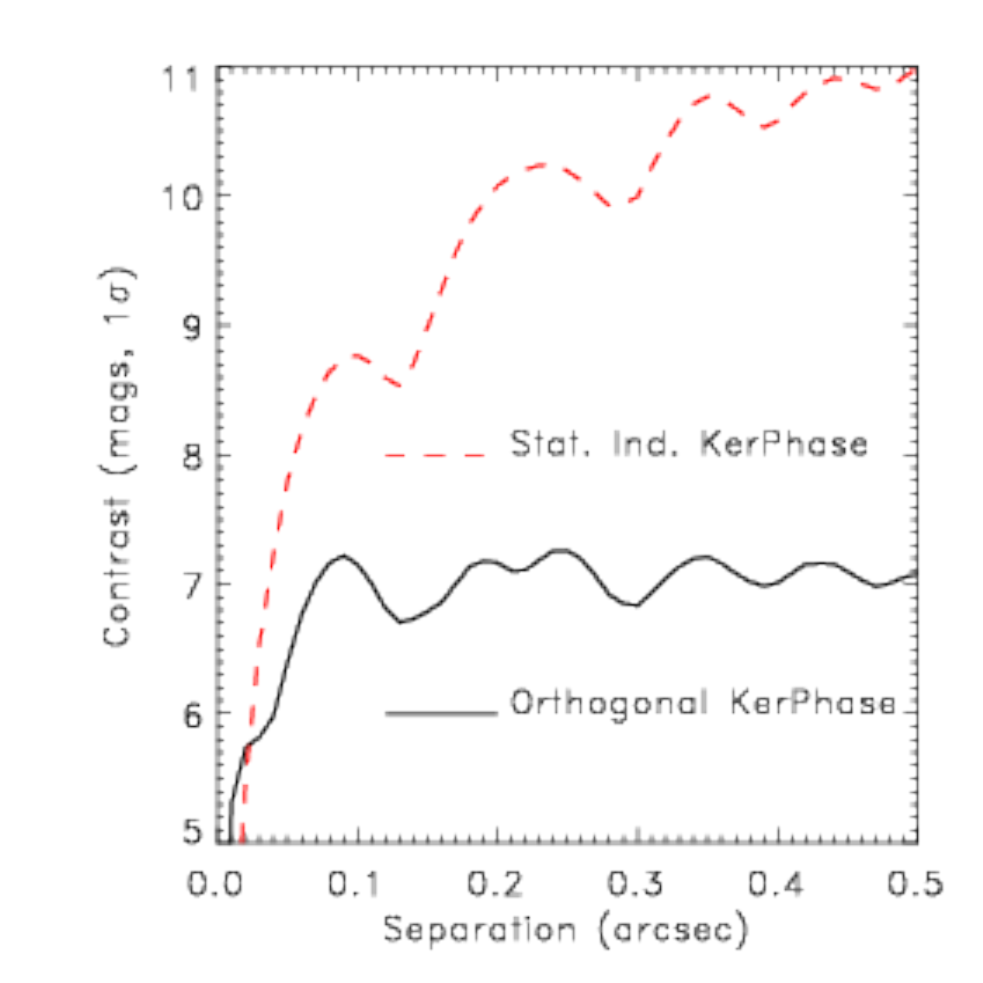}
\caption{The effect of photon-noise on Kernel-phase detections, based on a simulated photon-limited image with $10^6$ photons taken with the unobstructed Keck telescope in the Lp filter. The decreased number of photons far from the PSF core means that Kernel-phases sensitive to these spatial locations have smaller errors, increasing the achievable contrast. Although the Kernel-phases in
each situation are equivalent, the uncertainties are not equivalent, and would require a full covariance matrix in the case of the orthogonal kernel-phase.}
%statistically independent Kernel-phase. Two examples of a single Kernel-phase is shown as the mapping of the kernel-phase to the image-plane by adding together sinusoids with amplitudes equal to the kernel-phase weights. Where errors are dominated by photon-noise, these image-plane functions are spatially localised, and kernel-phases localised further from the PSF core have smaller variances.}
\label{figKernelPhotonNoise}
\end{figure}

\subsection{Statistically-Independent Kernel Phase}

Following from Section~\ref{kerphaseIntro} we will define the matrix that transforms the Fourier phase vector $\Phi$ to the vector of kernel-phases $\bs{K}_o$. This is an $N_K$ by $N_F$ matrix, where $N_K$ is the number of Kernel-phases and $N_F$ is the number of Fourier phases. The subscript $o$ indicates that this matrix produces an orthonormal set of phase linear combinations. We can 
compute the sample covariance matrix of kernel phases $\mathbf{C}_K$ either directly or from the sample covariance matrix of Fourier phases $\bs{C}$. This matrix can be diagonalised by the finite-dimensional spectral theorem:

\begin{equation}
 \mathbf{S}^T \cdot \mathbf{D} \cdot \mathbf{S} = \mathbf{C}_K = \mathbf{K}_o \cdot  \mathbf{C} \cdot \mathbf{K}_o^T.
 \label{eqnSDefn}
\end{equation}

The matrix $S$ is then a unitary matrix which allows us to construct a set of statistically independent kernel phases based on a new kernel-phase operator $\mathbf{K}_S$:

\begin{equation}
\theta_S = \mathbf{K}_S \cdot \Phi = \mathbf{S} \cdot \mathbf{K}_o \cdot \Phi.
\end{equation}

As an example of the utility of this approach, I have simulated the effects of photon-noise on Kernel-phase contrast limits, as shown in Figure~\ref{figKernelPhotonNoise}. The contrast standard deviation was estimated by first estimating the standard deviation of each Kernel-phase (i.e. neglecting covariances), forming a vector $\bs{\sigma}(\theta)$, then computing the contrast error using standard formulae for weighted averages:

\begin{align}
\theta_m &= \mathbf{K} \cdot \Phi_m \\
%m_\theta &=& \mathbf{K} \cdot m_\Phi \\
\sigma_c^2 &= 1 / \Sigma \frac{\theta_{m,k}^2}{\sigma^2_k(\theta)}
\end{align}

Here $\Phi_m$ is the model phase divided by the contrast in the high-conrast limit, e.g. for a 100:1 brightness ratio companion, the phase would be approximated well by 0.01$\Phi_m$. It is clear that the contrast achieved by considering statistically independent 
kernel-phases defined by $\bs{K}_S$ is superior to the contrast achieved by orthonormal kernel-phases defined by $\bs{K}_o$, for companions away from the PSF core.

\section{Calibration Strategies}
\label{sectCalibration}

For the situation where phase errors are mostly random, calibration is not required. This has been the case for faint 
aperture-mask observations with a laser-guide star system, where obtaining calibration observations has a very significant observing time cost \citep[e.g.][]{Dupuy09}. When static phase errors dominate and random errors are larger than calibration errors, only a single suitable calibrator observation is required. 
A more typical situation in sparse aperture masking has been where random errors are small compared to calibration errors, 
and the choice and weight assigned to calibrator observations is critical in achieving the lowest possible model fit residuals and the highest contrasts. In this regime there is an obvious danger -- where calibrators are chosen to minimise the calibrated kernel-phase, this biases the kernel-phase away from a detection, and may result in deeper contrast limits being quoted for a non-detection than is justified by the data. This problem is also in common with the LOCI algorithm \citep{Lafreniere07b}.

\subsection{Nearest Neighbour Calibration}
\label{sectNearestNeighbour}

The simplest calibration technique is to subtract the kernel-phases from a calibrator observed closest to the target in time or space. A small extension to this technique \citep[e.g.][]{Evans12}, is to use the average of several calibrators observed nearby in time, rejecting outlier calibrator observations. Outliers are most easily rejected by looking for calibrators that when used to calibrate the target, give spuriously large closure-phases. For $N_c$ calibrators, this amounts to calibrator weightings $\{a_k\}_{k=1}^{N_c}$ where each $a_k$ is either 0 or $1/N_u$, with $N_u$ the number of calibrators used. There are however, several weaknesses to this technique:

\begin{enumerate}
\item With small numbers of calibrator observations, it is difficult to avoid subjectivity in the choice to reject particular calibrators.
\item For particularly noisy calibrator observations and small systematic kernel phases, this process only adds noise.
\item All calibrators are weighted evenly, when the optimal weighting of individual calibrators may even be negative.
\item Any astrophysical structure in calibrators, e.g. undetected faint companions, contributes to any signal in final calibrated data.
\end{enumerate}

\begin{figure}
\includegraphics[scale=0.4]{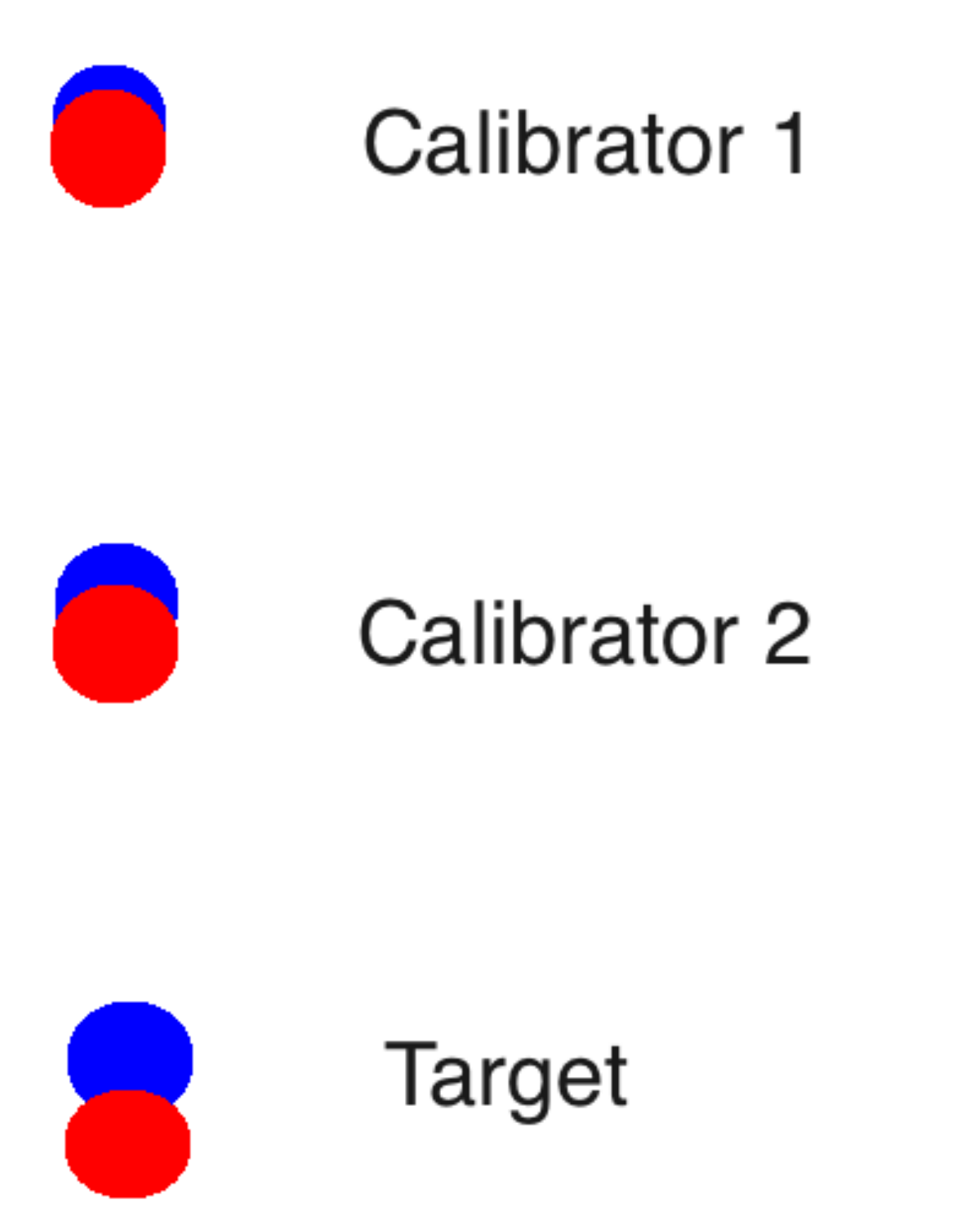}
\caption{An illustration of a situation where negative weighting of a calibrator may be optimal. Dispersion (illustrated by the red and blue circles) causes systematic kernel phases such that the kernel phases of Calibrator 2 ($\phi_{C2}$) is the average of kernel-phases of the Target ($\phi_T$) and Calibrator 1 ($\phi_{C1}$). The best estimate of the kernel-phases caused by dispersion for the Target is then $2\phi_{C2} - \phi_{C1}$.}
\label{figNegCalibrator}
\end{figure}

The third point may not be obvious, and is illustrated in Figure~\ref{figNegCalibrator}. Whenever calibrators are all on one side of the calibrator in some space, then optimal calibration may extrapolate past the position of the calibrators to the target. This space may be real (such as zenith distance which produces non-zero kernel phases due to dispersion) or a one dimensional parameterisation of a hidden variable describing a time-variable aberration. This approach is similar to the potentially negative weighting of astrometric reference stars in precision astrometry \citep{Lazorenko06}.

\subsection{Optimised Calibrator Weighting}
%\subsection{The Globally Optimised Phase Inference (GOPI) technique}
\label{sectGOPI}

We will now proceed to define a more optimal set of calibrator weightings $\{a_k\}_{k=1}^{N_c}$.  
This set of calibrator weightings must minimise the residual closure-phases after fitting a model, 
without significantly biasing the model fit. In this section, we will describe this process as applied in \citet{Kraus12}, 
where the starting point is closure-phases rather than kernel-phases.

Following Appendix~A of \citet{Kraus08}, we begin by considering the
closure-phases only on a subspace spanned by the $N_{\rm ind}$
linearly independent set of closure-phases. Furthermore, we construct a
basis vector set on this subspace such that the closure-phase
covariance matrix is diagonal (or nearly so) when projected on to it. To see how this
is done, first note how closure-phases can be constructed linearly from phases:

\begin{equation}
 \boldsymbol{\phi_{\rm cp}} = \bs{T} \boldsymbol{\theta_p}
\end{equation}

The matrix $\bs{T}\bs{T}^t$ then projects any set of closure-phases onto the set
spanned by the linearly independent set of closure-phases. This matrix
can be diagonalised 
$\bs{T}\bs{T}^t = \bs{U}_1^t\bs{D}_1\bs{U}_1$ 
by a diagonal matrix $\bs{D}_1$
and a unitary matrix $\bs{U}_1$. The eigenvalues on the diagonal of  $\bs{D}_1$
are either 0 or 1. By considering only the non-zero eigenvectors of
$\bs{D}_1$, we can write:

\begin{equation} 
\bs{T}\bs{T}^t = \bs{P}_1^t \bs{P}_1  \label{eqnCPProjection}
\end{equation}

for an $N_{\rm ind} \times
N_{\rm cp}$ projection matrix $\bs{P}_1$. $\bs{P}_1$ projects onto a subspace
$\mathbb{S}$ spanned by an orthonormal
set of linear combinations of closure-phases.

Next, given a closure-phase covariance matrix $\bs{C}_{\rm cp}$, we can modify the
projection matrix so that it projects onto a set of basis vectors for
$\mathbb{S}$ with a diagonal covariance matrix. To accomplish this, we
diagonalise the projection of $\bs{C}_{\rm cp}$:

\begin{equation}
 \bs{P}_1 \bs{C}_{\rm cp} \bs{P}_1^t = \bs{U}_2^t \bs{D}_2 \bs{U}_2.
\end{equation}

Then our new matrix $\bs{P}_2 = \bs{U}_2 \bs{P}_1$ is a projection matrix onto
$\mathbb{S}$ satisfying:

\begin{equation}
 \bs{P}_2 \bs{C}_{\rm cp} \bs{P}_2^t = \bs{D}_2,
\end{equation}

Representing the data in this way enables, for example, the
construction of $\chi^2$ variables that can be computed by the sum
over variance-normalised square deviates of
a set of independent data, without the explicit use of covariance
matrices. A potential problem with this approach is that the
sample covariance matrix estimated from the data has a rank equal to 
min$(N_{\rm ind}, N_{\rm fr}-1)$, where
$N_{\rm fr}$ is the number of data frames. Taken at face
value, with $N_{\rm fr} < N_{\rm ind}$, this process unreasonably restricts
the closure-phases of a model of the target to lie on a very limited subspace in the space spanned the
observed departures from the mean closure-phase. For this reason, we
take $\bs{C}_{\rm cp}$ above to be the weighted mean sample covariance matrix of all
target and calibrator observations weighted by the inverse of the trace of each
sample covariance matrix. We form the estimated errors of the target by:

\begin{equation}
 \bs{P}_2 \bs{C}_t \bs{P}_2^t = \bs{D}_2'
\end{equation}

Our data and errors are then transformed to a set of kernel-phases $\bs{x}$:

\begin{align}
 \bs{x} &= \bs{P}_2 \bs{\phi_{\rm cp}}\\
 \bs{\sigma^2(x)} &= {\rm diag}(\bs{D}'_2) + \Delta^2. \label{eqnDelta}
\end{align}

The non-diagonal terms of $\bs{D}'_2$ are ignored, and any values on the
diagonal less than the median are set to the median. This is a crude
method to ensure our statistics are reasonably robust, without
resorting to studentizing a multidimensional distribution. An alternative to this approach 
might be a bootstrapping
technique, however in this case there is no obvious way to estimate
the $a_k$ variables below or to account for the error in their
estimation. The additional uncertainty $\Delta^2$ accounts for calibration errors, 
to be further defined below.

%When considering binary fits at high contrast (i.e. a contrast ratio
%secondary/primary of $c \ll 1$), we make the
%approximation:
%
%\begin{eqnarray}
% \boldsymbol{\phi_b}(\rho, \theta, c) = c \boldsymbol{\phi_n}(\rho, \theta),
%\end{eqnarray}
%
%where the binary has a separation $\rho$, a position angle $\theta$
%and a contrast (secondary over primary) of $c$. This results in two
%new obvious subspaces, $\mathbb{B}$ with a projection matrix
%$\boldsymbol{\hat{\theta}\hat{\theta}^t}$, and $\mathbb{B}_\perp$ with a
%projection matrix:
%
%\begin{equation}
%  T T^t - \boldsymbol{\hat{\theta}\hat{\theta}^t} = P_3^t P_3.
%\end{equation}
%
%The matrix $P_3$ is constructed in a similar way to $P_2$ above, and
%has dimensions $N_{\rm ind}-1 \times N_{\rm cp}$. In a similar way to
%vectors $\boldsymbol{y}$, we define vectors
%$\boldsymbol{x}=P_3\boldsymbol{\phi_{\rm cp}}$. 
%
The next step is to find an optimal linear combination of weights
$\{a_k\}_{k=1}^{N_c}$, where $N_c$ is the number of possible calibrators. 
By {\em optimal}, we mean that we want to maximise the likelihood function for 
$\{a_k\}$ based on a null-model for calibrated kernel-phases $\bs{x}_c$:

% *** Originally, I had \hat{\boldsymbol{x_c}} , because it is really an estimator. Maybe confusing? ***
\begin{align}
\bs{x_c} &= \bs{x_t} - \Sigma_{k=1}^{N_c} a_k \bs{x_k} \\
%L(\{a_k\}) &=& \exp(-\hat{\bs{x_c}}^T \bs{C}_t^{-1} \hat{\boldsymbol{x_c}} /2 ) \pi(\{a_k\}) ,
L(\{a_k\}) &= \exp(-\Sigma_i \frac{x_{c,i}^2}{2\sigma_i^2 (\bs{x}_t)} ) \pi(\{a_k\}) , \label{eqnLikelihood}
\end{align}

where we have explicitly subscripted $\bs{x_c}$ with $i$ and where $\pi(\{a_k\})$ is a Bayesian prior distribution for $\{a_k\}$. 
The use of a restrictive prior as a regulariser 
is essential where there are many calibrators in use, because if $N_c > N_{\rm ind}$ and there is a
random error component, then there almost surely exists an $\{a_k\}$ such that $\bs{x_c} =0$, subtracting
any real astrophysical signal. The prior chosen in 
\citet{Kraus12}\footnote{This equation as presented in Equation 1 of \citet{Kraus12} was potentially confusing, because the division 
 $\frac{(\cdot)}{(\cdot)}$ was element-by-element division, and the vector $l^2$-norm $| \cdot |$ was used without being explicitly described.} 
 was:

\begin{equation}
 \pi( a_k ) = \exp \left (-\frac{a_k^2}{2} \Sigma_i \frac{\sigma_i^2(\boldsymbol{x_k})}{\sigma_i^2(\boldsymbol{x_t})} \right),
\end{equation}
 
where $\sigma_i^2(\boldsymbol{x})$ is the variance of the $i$-th component of $\boldsymbol{x}$. This is certainly not the only 
choice of such a prior, but it does have the essential feature of preferring calibrator weights of zero, and also of reducing the weighting of calibrators with large internal sample variances.
 
Once an optimal set of weights $\{ a_k \}$ has been found by maximising the likelihood function, the uncertainty on the 
calibrated kernel-phases $\bs{x_c}$ is given by:

\begin{equation}
 \sigma_i^2(\bs{x}_c) = \sigma_i^2(\bs{x}_t) + \Sigma a_k^2 \sigma_i^2(\bs{x}_k).
\end{equation}
 
Note that this neglects any uncertainty in estimating the $\{ a_k \}$. 

Finally, the calibrator observations $\{ \bs{x}_k \}$ do not necessarily span the space of the hidden parameters causing
non-zero point-source kernel-phases. For this reason, the additional ``calibration error'' term $\Delta^2$ in Equation 
\ref{eqnDelta} was iteratively added so that the reduced $\chi^2$ for the null-model was 1.0, i.e.:

\begin{equation}
 \chi^2_r = \frac{1}{N_{\rm ind}} \Sigma_i \frac{x_{c,i}^2}{\sigma^2_i(\bs{x_c})}= 1.0.
\end{equation}

In approximately half of the data sets tested in the work leading up to \citet{Kraus12}, no calibration error $\Delta^2$ was needed.
With values of the calibrated kernel-phases $\bs{x}_c$ and their errors
$\bs{\sigma}(\bs{x}_c)$ so computed, a model such as a bright star plus faint companion or 
a more complex image can be fit using least-squares. This is, however, a biased fit just like the LOCI 
technique \citep{Lafreniere07b}, 
because the process of computing the weights $\{ a_k \}$ partly removes the binary signal, due to the null model
for kernel-phase in Equation~\ref{eqnLikelihood}. For this reason, in \citet{Kraus12}, final values of model parameters 
were computed after re-computing the $\{ a_k \}$ with the best fit model subtracted iteratively from the $\bs{x}_c$.
 
\subsection{Restricted Kernel Phase (POISE)}
\label{sectPOISE}

An alternative to the complexity of the calibration strategy in the previous section is to ignore the kernel-phases that require calibration, i.e. those kernel phases that are most affected by systematic errors. This is similar to choosing a prior in
Equation~\ref{eqnLikelihood} so that the calibrator is ignored for some kernel-phases ($\pi(a_k) = \delta(0))$ and left
uniform for other kernel-phases, so that both calibration errors and astrophysical signal are subtracted. The difference between 
this and the technique described in this section is that only the restricted set of kernel-phases where calibration is not required
is used for subsequent analysis. We will call these restricted observables the Phase Observationally Independent of Systematic Errors ({\small POISE}) observables. This technique is very similar to the technique of ignoring dominant Karhunen-Lo\`eve eigenimages  as a means of calibrating more wide-field point-spread functions \citep{Soummer12}

Following Equation~\ref{eqnCPProjection}, we find a set of kernel-phases $\bs{y_k}$ for each image $k$ by a projection of the Fourier phases $\bs{\theta_p}$:

\begin{equation}
 \bs{y_k} = \bs{S}_c \bs{\theta}_p
\end{equation}

for general Kernel-phase, remembering that:

\begin{equation}
\bs{\theta}_p =  \bs{P}_1 \bs{\phi}_{\rm cp} = \bs{P}_1 \bs{T} \bs{\theta}_p
\end{equation}

for aperture-masking. The matrix $\bs{S}_c$ is formed in a similar way to Equation~\ref{eqnSDefn}, using the matrix $\bs{X} = \{ x_k\}$ of calibrator observations, which is an ( $N_K$ by $N_C$) matrix, with $N_C$ the total number of calibrator frames:

\begin{equation}
\bs{S}_c^T \cdot \bs{D} \cdot \bs{S}_c = \bs{X} \cdot \bs{X}^T.
\end{equation}

This definition is almost the same as taking diagonalizing the covariance matrix, except that we do not subtract the mean kernel-phases from the $\bs{x_k}$.

The calibrator kernel-phases on this new subspace $\bs{y_k}$ with zero covariances is are naturally subdivided into image sets $C_j$ for each PSF calibrator observation $j$. Within each image set, uncertainties are dominated by random errors, but between image sets, there is a combination of random and calibration errors. We consider the sample variance for kernel-phase $i$ computed over all images $k$ as {\em systematic} if:

\begin{equation}
 \delta_i^2 = s^2_i ( \{ \bs{y}_k \forall k \}) - s^2_i (\{ \bs{y}_k : k \in C_j\})  > 0
 \label{eqnSystematic}
\end{equation}

for all calibrator image sets $j$. 
%An easier way to determine if a kernel-phase is systematic is by determining if the deviation of kernel-phase $\langle y_k \rangle_{k \in C_j}$ averaged over each calibrator observation.
In the {\em POISE} technique, we simply compute the systematic error components $\delta^2_i$ for each kernel-phase $i$, and:

\begin{enumerate}
\item Ignore kernel-phases $y_i$ whenever
\begin{equation}
   \delta_i^2 >  \beta~\langle s_i^2(\{ y_k : k \in C_j \}) \rangle_j.
 \label{eqnBeta}
 \end{equation}  
A typical value for $\beta$ is 1, which rejects approximately 1 to 3 out of 28 kernel-phases for 9-hole Keck aperture-masking data.
\item Add $\delta_i^2$ to each target observation's uncertainty estimate for the remaining kernel-phases $i$.
\end{enumerate}

This means that the process of calibration is completely independent of the target, which was not the case in Section~\ref{sectGOPI}, because in that technique calibrator weights were chosen to minimise the calibrated target kernel phases. The technique requires at least 3 calibrator image sets to differ significantly from simpler calibration techniques.

As an example of the use of this technique, we consider the data set used in the November 2010 K' sparse aperture-mask 
observations of the LkCa~15 system \citep{Kraus12}. This data set consisted of 13 calibrator image sets of 12 images each, 
and 12 target image sets of 12 images each, all taken in good (0.6") seeing. This is an ideal data set, especially given that all 
calibrators had previous sparse aperture mask observations and were known to be single stars, and observations were 
continuous over a time period of 3.5 hours, with target and calibrator observations interspersed. This is also  the highest contrast 
detection published in the literature so far, which is the K-band detection of structure modelled as three compact sources around 
the star, with details reproduced in Table~\ref{tblLkCa15TripleFit}. Although much higher contrast is possible for brighter stars, 
especially when extreme adaptive optics may enable negligible piston phase errors, at $V\sim12$ this is roughly 
the brightest star of its class -- no known $<$5\,Myr  solar-mass star is in any association closer than Taurus.
 
When applying the POISE algorithm to this data set with a $\beta$ value of 1.0 in Equation~\ref{eqnBeta}, only 1 of the 28 
kernel-phases are removed as ``systematic'' by the calibrator observations, meaning that 96\% of the closure-phase information is 
retained. A three point-source fit to these restricted kernel-phases had a reduced $\chi^2$ value 
of 0.92, as shown in Table~\ref{tblLkCa15TripleFit}. With a reduction of $\beta$ to 0.25, 4 kernel-phases are removed as ``systematic'', the 
reduced $\chi^2$ becomes 1.00 but no fitted parameters change by even 1-$\sigma$. In addition, the 
variance of the mean for 50\% of the image-set kernel-phases are dominated by random errors, and not the $\delta_i^2$ values from 
Equation~\ref{eqnSystematic}.
This means that quasi-static spatial aberrations 
in this case do not significantly limit the signal-to-noise in the final image. For this kind of observation, spatially-filtering the input 
wavefront \citep[e.g.][]{Huby12,Jovanovic12} could not 
significantly improve the achievable calibration-limited contrast. The random errors of $\sim$0.5 degrees in each 240\,s image set are also 
consistent with temporal phase piston errors, which would not be improved by spatial filtering. 
This argument of course falls over for brighter targets (i.e. generally higher-mass or closer and older targets) where exiting adaptive optics systems 
perform much better, and extreme AO is possible. In these situations, $\sigma_\varphi$ in Equation~\ref{eqnTemporalErrors} can be smaller 
than 0.3 radians, $f_c$ can exceed 100\,Hz and spatial filtering may become essential at the $\sim$10\,magnitude contrast range enabled by 
this improved AO performance.

\subsection{Imaging with {\small POISE}}

For sufficiently complex sources, model-fitting is replaced with imaging. In general, imaging from kernel-phases alone is 
computationally intensive because of the nonlinear relationship between the image-plane and Fourier phase. However, 
in the high contrast regime,where interferometric visibility amplitudes are unity within errors, we can approximate the 
Fourier transform $F(\bs{u})$ of an image $I(\bs{x})$ normalised to a total flux of unity as:

\begin{align}
 F(\bs{u}) &\approx 1 + i \int \sin(2 \pi \bs{u} \cdot \bs{x})I(\bs{x}) d\bs{x}.
\end{align}

In turn, the phase $\Phi$ becomes:

\begin{align}
 \Phi(\bs{u}) &\approx  \int \sin(2 \pi \bs{u} \cdot \bs{x})I(\bs{x}) d\bs{x}.
 \label{eqnLinearPhi}
 \end{align}
 
We can consider the image to be made of discrete pixel values arranged in a vector $\bs{p} = \{ p_j \}$, so that the integral 
in Equation~\ref{eqnLinearPhi} becomes a sum, and the values of Fourier phases $\bs{\phi}$ and Kernel-phases $\bs{\theta}$ 
are represented by matrix multiplication:
 
 \begin{align}
      \bs{\Phi}   &\approx \bs{M} \cdot \bs{p} \\
      \bs{\theta} &\approx \bs{K} \cdot \bs{M} \cdot \bs{p} \nonumber \\
                         &\approx \bs{A} \cdot \bs{p}.
\end{align}

This linear approximation to imaging means that minimising kernel-phase $\chi^2$ subject to a differentiable
regulariser can be rapidly computed using a gradient descent method. An example of such a regulariser is the 
Maximum Entropy regularizer \citep[e.g.][]{Narayan86}:

\begin{align}
 S = -\Sigma_j p_j \ln(p_j/q_j),
\end{align}

for some prior image $\bs q$, often taken to be a uniform image in some finite field of view and zero elsewhere. The problem of Maximum Entropy image construction is then simply a problem of minimising the sum of the $\chi^2$ value and the regulariser:

\begin{align}
 \bs{p}_{\rm MaxEnt} = {\rm arg\,min}_{\bs {p}} \left\{ \Sigma_i \frac{(\theta_i - \bs{A} \cdot \bs{p})^2}{\sigma^2_i}  + \alpha \Sigma_j p_j \ln(p_j/q_j) \right\}.
\end{align}

\begin{figure}
\includegraphics[width=1.05\columnwidth]{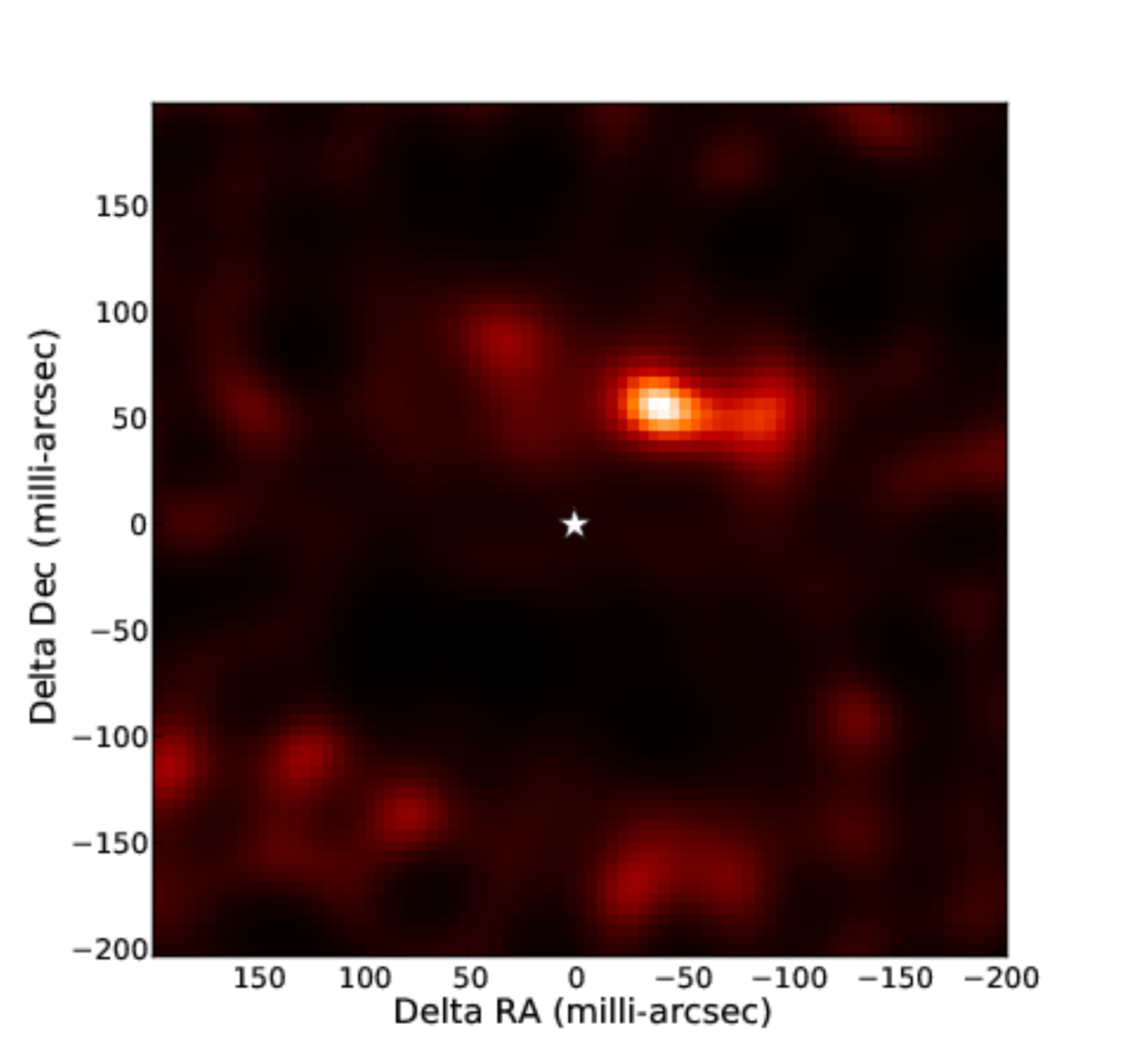}
\caption{A imaging fit to the 2010 November data set of LkCa~15, originally published in \citet{Kraus12}.
A uniform prior was used which had
a total flux of 2\% of the image flux, and the final fit at a reduced $\chi^2$ of 1.0 contained 1\% of the image flux, with
the remaining 99\% contained within the point source star at the image centre.}
\label{figLkCa15_K10}
\end{figure}

The value of $\alpha$ is typically chosen so that the final image has a reduced $\chi^2$ value of 1.0\footnote{Image reconstruction code in the {\tt python} language using this regulariser can be found at 
{\tt http://code.google.com/p/pysco}, the repository where all code in this paper is intended to go after translation to {\tt python}.}.
%RRR New sentences
The to see the result of this approach to imaging, we will again use the K' data set from \citet{Kraus12}. In
that original paper, the optimised calibrator weighting scheme (see Section~\ref{sectGOPI}) enabled the MACIM algorithm \citep{Ireland06b} to be used to create images directly from the closure-phases via an {\small OIFITS} input file. This approach ignored correlations between closure-phases.  The image created directly by fitting to kernel-phases imaging with the Maximum Entropy regulariser can be seen in Figure~\ref{figLkCa15_K10}, where the resolved structures contain 1\% of the total system flux and the reduced $\chi^2$ of the image is 1.0. Note that arbitrary point-symmetric 
flux could be added to this image and it would still fit the Kernel-phases. A weakness of imaging from kernel-phases alone is that point-symmetric flux 
added to a bright central point source does not produce any phase information.

The image in Figure~\ref{figLkCa15_K10} is cosmetically at least as good cosmetically as that shown in \citet{Kraus12}, but comes with the
significant benefit that the calibration process does not directly affect the image: the {\small POISE} observables are independent of 
the calibrator observations. 

\begin{table}
\caption{A comparison between a three additional point-source fit to
2010 November K' sparse aperture mask data using linear combinations of
calibrator observations \citep[][KI12]{Kraus12} and using the {\small POISE}
observables. Parameters are separation ($\rho$), position angle ($\theta$) 
and magnitude difference with respect to the primary ($\Delta m$). 
When adding uncertainties in quadrature, differences 
are always consistent within 2-$\sigma$ and in 7 out 
of 9 cases within 1-$\sigma$.}
\begin{tabular}{lrr}
Parameter & KI12 & {\small POISE} \\
\hline
$\rho_1$ (mas) &67.0$\pm$3.2 & 65.1$\pm$3.1 \\
$\theta_1$ (deg) & 12.3$\pm$2.8& 10.9$\pm$2.9 \\
$\Delta m_1$ & 7.40$\pm$0.19& 6.89$\pm$0.18 \\
$\rho_2$ (mas) & 64.4$\pm$1.5 & 62.6$\pm$1.9 \\
$\theta_2$ (deg) & 334.8$\pm$1.5 & 333.4$\pm$2.5 \\
$\Delta m_2$ & 6.59$\pm$0.09 & 6.36$\pm$0.11 \\
$\rho_3$ (mas) & 82.5$\pm$2.4 & 78.0$\pm$4.1 \\
$\theta_3$ (deg) & 302.3$\pm$1.5 & 302.3$\pm$2.8 \\
$\Delta m_3$ & 7.06$\pm$0.12 & 7.02$\pm$0.18 \\
\end{tabular}
\label{tblLkCa15TripleFit}
\end{table}

%\begin{table}
%\caption{The same as Table~\ref{tblLkCa15TripleFit} for a 2-component
%fit to the 2010 November L' band data.}
%\begin{tabular}{lrr}
%Parameter & KI12 & {\small POISE} \\
%\hline
%$\rho_1$ & 93.8$\pm$3.4 & \\
%$\theta_1$ & 356.7$\pm$2.4 & \\
%$\Delta m_1$ & 5.33$\pm$0.15& \\
%$\rho_2$ & 84.1$\pm$2.8 & \\
%$\theta_2$ & 310.0$\pm$2.0 & \\
%$\Delta m_2$ & 5.28$\pm$0.08 & \\
%\end{tabular}
%\label{tblLkCa15Lfit}
%\end{table}

%Each Kernel-phase has its own image in the high-contrast regime. This means that:
% \theta_m = M \cdot I_{m,k}
% chi^2({I_k}) = \Sigma (\theta_d - M \cdot I_{m,k})^2/\sigma_d^2
% d\chi^2 d I_k = 2 \Sigma_j M_{j,k} (\theta_d - M \cdot I_{m,k})/\sigma_d^2
%É which means we could just do gradient descent. 
%There must be a way to hard-code this, because for a direction g,  a starting point x and s distance t,
%the matrix product is just M \cdot (x + gt) = M \cdot x + t M \cdot g, meaning that only one matrix
%multiplication is needed per line search. Nice and fast.
%
%NB there seems to be no way to make a "dirty map" which like in Radio is equally positive and negative. 
%The problem with this is that there is no convolution operator so standard codes won't work.

\section{Conclusions}
\label{sectConclusions}

Aperture-mask interferometry has proven to be a powerful technique to recover high contrast (up to $\sim$8\, magnitudes at 1$\sigma$), asymmetric information at the diffraction limit ($\sim$ 0.5--$5 \lambda/D$) of large telescopes. The reason for this success is the ability for closure-phase, a kind of kernel-phase, to give an observable largely independent of time-variable aberrations. I have described many of the key sources of phase errors in this technique, as well as several strategies for mitigating them. Of note is the Phase Observationally Independent of Systematic Errors ({\small POISE}) observables, which are a subset of all possible linear combinations of closure-phases. Observations of calibrator stars inform which linear combinations of phases constitute the {\small POISE} observables, but the analysis of the target observations is performed quite independently of the calibrator observations, leading to a more robust calibration method.

The generalisation of the aperture-mask technique to full pupil images shows great promise in the form of the full pupil kernel-phase observables. Simulations show that pupil-plane phase errors higher than third-order affect full pupil kernel-phase more than aperture-mask kernel-phase, meaning that full-pupil kernel phase will likely be restricted to moderately high Strehl observations.

%RRR Changes (JGR)
The analysis presented here has implicitly involved only a monochromatic PSF from an imaging system. Although the effect of dispersion was discussed and the POISE calibration technique ameliorates the effects of dispersion, a mathematical framework to clearly predict the effects of dispersion on kernel-phase was not developed. A future study of the effect of very broad bandwidths is needed. More importantly, an extension of this technique to work for the simultaneous wavelength-dispersed images formed by an integral field unit could be very powerful. The scaling of PSF with wavelength as a speckle suppression technique could be equally-well applied to observables in the Fourier domain as it has been in image-plane analyses.

\section*{Acknowledgments}
M.I. would like to acknowledge many helpful conversations with and encouragement from 
a large number of people over the past 10 years as these ideas developed and have been 
tested in various contexts, in particular 
Jean-Philippe Berger, Adam Kraus, Shri Kulkarni, Sylvestre Lacour, David Lafreni\'ere, 
James Lloyd,  Frantz Martinache, John Monnier, Laurent Puyeyo, 
J. Gordon Robertson, Anand Sivaramakrishnan and Peter Tuthill.
%RRR
The manuscript was also substantially improved following helpful comments
from an anonymous referee.

\appendix

\section{Third-Order Bispectrum Expansion}

We will begin by writing the combination of Equations~\ref{eqnThirdOrderV} and \ref{eqnBispect} explicitly:

\begin{align}
b_{ABC} &= (1 + i \av{(\pb - \pa)} - \frac{1}{2}\av{(\pb-\pa)^2} - \frac{i}{6}\av{(\pb-\pa)^3})  \nonumber \\
& \times (1 + i \av{(\pc - \pb)} - \frac{1}{2}\av{(\pc-\pb)^2} - \frac{i}{6}\av{(\pc-\pb)^3})  \nonumber \\
& \times (1 + i \av{(\pa - \pc)} - \frac{1}{2}\av{(\pa-\pc)^2} - \frac{i}{6}\av{(\pa-\pc)^3}) \label{eqnBispectFull}
\end{align}

The 0th order terms in the $\Delta \phi$s are trivially collected as 1, and the 1st order terms clearly cancel to give 0. The second order terms are:

\begin{align}
 \Re(b_{ABC})&\approx -\frac{1}{2}[ ~ \av{(\pb - \pa)^2} + \av{(\pc - \pb)^2} + \av{(\pa-\pc)^2} ~] \nonumber \\
 & - [~ \av{(\pb - \pa)} \cdot \av{(\pc - \pb)}+  \av{(\pc - \pb)}\cdot \av{(\pa - \pc)} \nonumber \\
 & +  \av{(\pa - \pc)}\cdot \av{(\pb - \pa)}~ ]
\end{align}

Moving from this equation to Equation~\ref{eqnRealBispect} requires the substitution of Equations~\ref{eqnPistonA} through 
\ref{eqnPistonC}, as well as a recognition of the following classes of trivial identities:

\begin{align}
\av{(\pb -\pa)} &= (\av{\pb} - \av{\pa}) \label{eqnIdentity1} \\
\av{(\pb' - \pa')} &= 0 \label{eqnIdentity2}
\end{align}

The 3rd order terms of Equation~\ref{eqnBispectFull} are collected (after minor simplification of the coefficient 1/2 terms) as:

\begin{align}
\Im(b_{ABC})&\approx -\frac{1}{6}[~ \av{(\pb - \pa)^3} + \av{(\pc - \pb)^3} + \av{(\pa-\pc)^3}~] \nonumber \\
 & +\frac{1}{2} [~  \av{(\pb - \pa)} \cdot \av{(\pb - \pa)^2} \nonumber \\
 & + \av{(\pc - \pb)} \cdot\av{(\pc - \pb)^2} + \av{(\pa-\pc)}\cdot\av{(\pa-\pc)^2}~] \nonumber \\
 &  - \av{(\pb - \pa)}\cdot \av{(\pc - \pb)}\cdot \av{(\pa - \pc)}.
\end{align}

Again, Equation~\ref{eqnImBispect} follows after substitution of Equations~\ref{eqnPistonA} through 
\ref{eqnPistonC} as well as applying trivial identities such as \ref{eqnIdentity1} and \ref{eqnIdentity2}.

\section{Temporal Phase Errors}

In applying Equation~\ref{eqnImBispect} to  temporal phase errors, we write the instantaneous values of $\pa$, $\pb$ and $\pc$ as random variables $X_A$, $X_B$ and $X_C$ respectively, which take a new random value at $N$ statistically independent time steps. We can then write:

\begin{align}
{\rm Var}(\phi_{\rm cp}) &= \frac{1}{36} {\rm Var}(\av{(\pb' - \pa')^3} + \av{(\pc' - \pb')^3} + \av{(\pa'-\pc')^3}) \\
 &\approx \frac{1}{36 N} {\rm Var}( (X_B - X_A)^3  \nonumber \\
 & + (X_C - X_B)^3 + (X_A - X_C)^3) \label{eqnApproxExpansion} \\
 &=  \frac{1}{4 N} {\rm Var}( X_A^2 X_B - X_A X_B^2 + X_B^2 X_C \nonumber \\
  &- X_B X_C^2 + X_A X_C^2 - X_A^2 X_C). \label{eqnExpansion} \\
 &= \frac{3 \sigma_\varphi^6}{N}. \label{eqnExpansionResult}
\end{align}

% The Variance expands by collecting terms of a common form as:
% 6 Var (X_A^2 X_B) + 12 Cov(X_A X_B^2, X_A^2 X_C) - 6 Cov(X_A X_B^2,X_A^2 X_B) - 6 Cov(X_A X_B^2,X_A X_C^2 ) - 6 Cov(X_A^2 X_B,X_A^2 X_C)
%... and where we have an E(A^4), this is 3\sigma_\varphi^4 from standard (Wikipedia) properties of a Gaussian.

Here Var represents the variance of a quantity, which in this special case of quantities of zero mean, is simply the expectation of the square. The approximately equals sign ($\approx$) in Equation~\ref{eqnApproxExpansion} is used because we are ignoring the piston subtraction, applicable only for $N>>1$ (and with an error of order $N^{-1}$). Each of the variables $X_A$, $X_B$ and $X_C$ are independent Gaussian variables with mean 0 and standard deviation $\sigma_\varphi$, so their moments are standard results, and the expectation of a product of their moments is simply the product of the expectation of their respective moments.  The variance on the right hand side of Equation~\ref{eqnExpansion} can be thus be simply but tediously evaluated as the sum over 36 mutual covariances to give a value of 12$\sigma_\varphi^6$. Finally, Equation~\ref{eqnTemporalErrors} follows directly from Equation~\ref{eqnExpansionResult}, noting that the number of independent phase samples $N = f_c T$.

%\bibliography{../mireland}

\begin{thebibliography}{25}
\expandafter\ifx\csname natexlab\endcsname\relax\def\natexlab#1{#1}\fi

\bibitem[{{Baldwin} {et~al}\mbox{.}(1986){Baldwin}, {Haniff}, {Mackay}, \&
  {Warner}}]{Baldwin86}
{Baldwin} J.~E., {Haniff} C.~A., {Mackay} C.~D., {Warner} P.~J., 1986, \nat,
  320, 595

\bibitem[{{Dupuy} {et~al}\mbox{.}(2009){Dupuy}, {Liu}, \& {Ireland}}]{Dupuy09}
{Dupuy} T.~J., {Liu} M.~C., {Ireland} M.~J., 2009, \apj, 699, 168

\bibitem[{{Evans} {et~al}\mbox{.}(2012){Evans}, {Ireland}, {Kraus},
  {Martinache}, {Stewart}, {Tuthill}, {Lacour}, {Carpenter}, \&
  {Hillenbrand}}]{Evans12}
{Evans} T.~M. {et~al.}, 2012, \apj, 744, 120

\bibitem[{{Fizeau}(1868)}]{Fizeau1868}
{Fizeau} H., 1868, C.R.Acad.Sci., 66, 932

\bibitem[{{Hinkley} {et~al}\mbox{.}(2011){Hinkley}, {Carpenter}, {Ireland}, \&
  {Kraus}}]{Hinkley11}
{Hinkley} S., {Carpenter} J.~M., {Ireland} M.~J., {Kraus} A.~L., 2011, \apjl,
  730, L21

\bibitem[{{Hofmann} \& {Weigelt}(1993)}]{Hofmann93}
{Hofmann} K.-H., {Weigelt} G., 1993, \aap, 278, 328

\bibitem[{{Huby} {et~al}\mbox{.}(2012){Huby}, {Perrin}, {Marchis}, {Lacour},
  {Kotani}, {Duch{\^e}ne}, {Choquet}, {Gates}, {Woillez}, {Lai}, {F{\'e}dou},
  {Collin}, {Chapron}, {Arslanyan}, \& {Burns}}]{Huby12}
{Huby} E. {et~al.}, 2012, \aap, 541, A55

\bibitem[{{Ireland} {et~al}\mbox{.}(2008){Ireland}, {M{\'e}rand}, {ten
  Brummelaar}, {Tuthill}, {Schaefer}, {Turner}, {Sturmann}, {Sturmann}, \&
  {McAlister}}]{Ireland08c}
{Ireland} M.~J. {et~al.}, 2008, in Society of Photo-Optical Instrumentation
  Engineers (SPIE) Conference Series, Vol. 7013, Society of Photo-Optical
  Instrumentation Engineers (SPIE) Conference Series

\bibitem[{{Ireland} {et~al}\mbox{.}(2006){Ireland}, {Monnier}, \&
  {Thureau}}]{Ireland06b}
{Ireland} M.~J., {Monnier} J.~D., {Thureau} N., 2006, in Society of
  Photo-Optical Instrumentation Engineers (SPIE) Conference Series, Vol. 6268,
  Society of Photo-Optical Instrumentation Engineers (SPIE) Conference Series

\bibitem[{{Jovanovic} {et~al}\mbox{.}(2012){Jovanovic}, {Tuthill}, {Norris},
  {Gross}, {Stewart}, {Charles}, {Lacour}, {Ams}, {Lawrence}, {Lehmann},
  {Niel}, {Robertson}, {Marshall}, {Ireland}, {Fuerbach}, \&
  {Withford}}]{Jovanovic12}
{Jovanovic} N. {et~al.}, 2012, \mnras, 427, 806

\bibitem[{{Kraus} \& {Ireland}(2012)}]{Kraus12}
{Kraus} A.~L., {Ireland} M.~J., 2012, \apj, 745, 5

\bibitem[{{Kraus} {et~al}\mbox{.}(2008){Kraus}, {Ireland}, {Martinache}, \&
  {Lloyd}}]{Kraus08}
{Kraus} A.~L., {Ireland} M.~J., {Martinache} F., {Lloyd} J.~P., 2008, \apj,
  679, 762

\bibitem[{{Kulkarni}(1989)}]{Kulkarni89}
{Kulkarni} S.~R., 1989, \aj, 98, 1112

\bibitem[{{Lacour} {et~al}\mbox{.}(2011){Lacour}, {Tuthill}, {Amico},
  {Ireland}, {Ehrenreich}, {Huelamo}, \& {Lagrange}}]{Lacour11}
{Lacour} S., {Tuthill} P., {Amico} P., {Ireland} M., {Ehrenreich} D., {Huelamo}
  N., {Lagrange} A.-M., 2011, \aap, 532, A72

\bibitem[{{Lafreni{\`e}re} {et~al}\mbox{.}(2007){Lafreni{\`e}re}, {Marois},
  {Doyon}, {Nadeau}, \& {Artigau}}]{Lafreniere07b}
{Lafreni{\`e}re} D., {Marois} C., {Doyon} R., {Nadeau} D., {Artigau} {\'E}.,
  2007, \apj, 660, 770

\bibitem[{{Lazorenko}(2006)}]{Lazorenko06}
{Lazorenko} P.~F., 2006, \aap, 449, 1271

\bibitem[{{Le Bouquin} \& {Absil}(2012)}]{leBouquin12}
{Le Bouquin} J.-B., {Absil} O., 2012, \aap, 541, A89

\bibitem[{{Lloyd} {et~al}\mbox{.}(2006){Lloyd}, {Martinache}, {Ireland},
  {Monnier}, {Pravdo}, {Shaklan}, \& {Tuthill}}]{Lloyd06}
{Lloyd} J.~P., {Martinache} F., {Ireland} M.~J., {Monnier} J.~D., {Pravdo}
  S.~H., {Shaklan} S.~B., {Tuthill} P.~G., 2006, \apjl, 650, L131

\bibitem[{{Martinache}(2010)}]{Martinache10}
{Martinache} F., 2010, \apj, 724, 464

\bibitem[{{Michelson}(1891)}]{Michelson1891}
{Michelson} A.~A., 1891, \nat, 45, 160

\bibitem[{{Monnier} {et~al}\mbox{.}(2006){Monnier}, {Pedretti}, {Thureau},
  {Berger}, {Millan-Gabet}, {ten Brummelaar}, {McAlister}, {Sturmann},
  {Sturmann}, {Muirhead}, {Tannirkulam}, {Webster}, \& {Zhao}}]{Monnier06}
{Monnier} J.~D. {et~al.}, 2006, in Society of Photo-Optical Instrumentation
  Engineers (SPIE) Conference Series, Vol. 6268, Society of Photo-Optical
  Instrumentation Engineers (SPIE) Conference Series

\bibitem[{{Narayan} \& {Nityananda}(1986)}]{Narayan86}
{Narayan} R., {Nityananda} R., 1986, \araa, 24, 127

\bibitem[{{Schwarzschild}(1896)}]{Schwarzschild1896}
{Schwarzschild} K., 1896, Astronomische Nachrichten, 139, 353

\bibitem[{{Soummer} {et~al}\mbox{.}(2012){Soummer}, {Pueyo}, \&
  {Larkin}}]{Soummer12}
{Soummer} R., {Pueyo} L., {Larkin} J., 2012, \apjl, 755, L28

\bibitem[{{Tuthill} {et~al}\mbox{.}(2000){Tuthill}, {Monnier}, {Danchi},
  {Wishnow}, \& {Haniff}}]{Tuthill00}
{Tuthill} P.~G., {Monnier} J.~D., {Danchi} W.~C., {Wishnow} E.~H., {Haniff}
  C.~A., 2000, \pasp, 112, 555

\end{thebibliography}
%\bibliographystyle{../cls/apj.bst}
%\bibliographystyle{mn2e}

\label{lastpage}

\end{document}